\documentclass[showkeys,preprint,showpacs,preprintnumbers, amsmath,amssymb]{revtex4-1}

\usepackage{multirow}

\usepackage{graphics}%
\usepackage{epsfig}%
\usepackage{subfigure}

\usepackage[latin1]{inputenc}
\usepackage{t1enc}

\usepackage{algorithmic}

\begin{document}

\title{Using graphics processing units to generate random numbers}

\author{Sami Hissoiny}
\email{sami.hissoiny@polymtl.ca}
\affiliation{École polytechnique de Montréal, Département de génie
informatique et génie logiciel, 2500 chemin de Polytechnique.
Montréal (Québec), CANADA H3T 1J4}
\author{Philippe Després}
\affiliation{Département de Physique, de Génie Physique et d'Optique, Université, Laval, Québec, Qc, Canada and Département de Radio-Oncologie and Centre de Recherche en Cancérologie de l'Université Laval, Centre Hospitalier Universitaire de Québec (CHUQ), Québec, Qc, Canada}
\author{Benoît Ozell}
\affiliation{École polytechnique de Montréal, Département de génie
informatique et génie logiciel, 2500 chemin de Polytechnique.
Montréal (Québec), CANADA H3T 1J4}

\begin{abstract} The future of high-performance computing is aligning itself towards the efficient use of
highly parallel computing environments. One application where the use of
massive parallelism comes instinctively is Monte Carlo simulations, where a
large number of independent events have to be simulated. At the core of the
Monte Carlo simulation lies the Random Number Generator (RNG). In this paper,
the massively parallel implementation of a collection of pseudo-random number
generators on a graphics processing unit (GPU) is presented. The results of the
GPU implementation, in terms of samples/s, effective bandwidth and operations
per second, are presented. A comparison with other implementations on different
hardware platforms, in terms of samples/s, power efficiency and cost-benefit,
is also presented. Random numbers generation throughput of up to
$\approx$18MSamples/s have been achieved on the graphics hardware used.
Efficient hardware utilization, in terms of operations per second, has reached
$\approx$98\% of the possible integer operation throughput.
\end{abstract}


\maketitle

\section{Introduction}
Over the years, the increase in computing power has been driven by the increase
in processor clock speed. Lately, however, the increase in computing power has
been achieved by the use of multi core CPUs (Central Processing Units) while
raw clock frequencies have evolved at a slower pace. Recent trends in
scientific computing point towards massively parallel computing devices to
handle much of the workload. One class of hardware that has always been
designed with parallelism in mind, due to the specialized task it accomplishes,
is the Graphics Processing Unit (GPU). The scientific computing community is
turning more and more towards this hardware platform as it is well suited for a
class of problem often encountered when the intended goal is to simulate real
phenomena: embarrassingly parallel computations. One family of applications
among this paradigm, Monte Carlo simulations, rely on high-quality
pseudo-random numbers. In this paper, we implement three algorithms for
generating random numbers: the multiply-with-carry (MWC), the XorShift, and the
KISS (keep it simple stupid). These implementations of random number generators
(RNG) will be analyzed in terms of their samples/s capability as well as their
ability to exploit the resources of the GPU. It should be noted that the RNG of
choice when long periods are wanted, the Mersenne Twister, has not been
implemented for reasons that will be explained lated.

Random number generators have been implemented on traditional CPUs as well as
more parallel computing oriented platforms such as GPUs~\cite{LANG09,JANO08}
and FPGAs~\cite{VINA07,THOM09}. A comparison to these implementations is also
presented.

\section{Background}
\subsection{Random number generators}
RNGs in general have a history of bad implementations, as pointed out by the
work of Park and Miller~\cite{PARK88}. For this reason, well established random
number generators will be used in this work, which is focused on their
implementation on GPUs. This work is based on the RNGs by George
Marsaglia~\cite{MARS03,MARS03a}. More precisely, a \emph{multiply-with-carry}
generator, a \emph{XorShift} generator and the \emph{KISS} generator have been
implemented.

RNGs are based on the execution of mathematical operations on an ensemble of variables (the RNG
state) to generate a sequence of numbers that will appear random; being generated through an
algorithm they are obviously completely deterministic. The sequence takes the form of:
\begin{equation}
x,f(x),f^2(x),f^3(x),...
\end{equation}
where x is the first value, or set of values, supplied to the RNG, called the seed, and $f^2(x)$ is equal
to $f(f(x))$. This recurrence will generate a seemingly random sequence until it has been called a
number of times equal to its period. After this point, the sequence starts over.

\subsubsection{\label{ssmwc}multiply-with-carry}
The multiply-with-carry (MWC) generator has been introduced by Marsaglia in
1991~\cite{MARS91}. The generator takes the form of:
\begin{equation}
x_{n+1}=(a*x_n+c_n) \bmod(b)
\end{equation}
where $a$ is the multiplier and $b$ the base. The carry, $c$, is defined by:
\begin{equation}
c_{n+1}=\lfloor \frac{(a*x_n+c_n)}{b} \rfloor.
\end{equation}
When using integer arithmetic on computers, it is advantageous to use a base such as $2^{16}$ or
$2^{32}$ in order to avoid the costly operations required to find the carry and replace them by very efficient shift
operations. The carry of the operation $a*x$ with base $b$ is $c=a*x>>b$ where $>>$ is the shift right ($shr$) operation. The modulo operation can also be transformed into an $ADD$ operation. Indeed, $a \bmod(2^{16})== a\&\&2^{16}$,

The choice of the multiplier $a$ is not arbitrary: the multiplier is chosen so that $ab-1$ is
safeprime. The period of the RNG with such a multiplier is on the order of $(ab-1)/2$.

This generator is characterized by a very short state vector of only one
element of $w$ bits where $w$ is usually 16 or 32. \emph{Lagged} MWCs can also
be developed. These \emph{lagged} have a longer state vectors which act as a
circular buffer of seeds for the MWC generator.

\subsubsection{ \label{secxorshift} XorShift}

The XorShift generator was introduced by Marsaglia~\cite{MARS03} and further
improved by Panneton~\cite{PANN09} and Brent~\cite{BREN04}. The improvements
proposed by Panneton will be used in this work. This RNG, as its name implies,
is based on the successive application of xorshift operations on the state
vector. The xorshift operation consists in replacing the $w$-bit block by a
bitwise xor (exclusive or) of the original block with a shifted copy of itself
by a position either to the right or to the left, where \mbox{0 < a < w} and
\mbox{w = 32 or 64}. These generators have periods equal to $2^{rw}-1$.

The generators are characterized by a number of xorshift operations on the
state vector to output one random number. The mathematics behind this are well
beyond the scope of this paper and are expertly explained by
Panneton~\cite{PANN09}. Of particular interest is that the RNG is parametrized
by $r$, the number of $w$-bit elements in the vector and $s$ the number of
xorshift operations. The sequence is defined by:
\begin{equation}
v_i=\sum_{j=1}^r\{A_jv_{i-j}\}
\end{equation}
where the $v_i$'s are $w$ bits long and form the state vector and the $A_j$'s are the product of $v_j$
xorshift matrices. There are $s$ non-zero $A_j$ matrices.

The generators are very fast since they only require bitwise and shift operations on unsigned integers,
which are generally some of the fastest instructions on a processor.

Panneton proposes a full period generator with $r=8$, $s=7$ resulting in a period of $2^{256}-1$. The
generator has the  following recurrence:

\begin{equation}
\begin{split}
v_n = (I + L^9)(I + L^{13})v_{n-1} + (I + L^7)v_{n-4}
\\+ (I +R^3)v_{n-5} + (I+R^{10})v_{n-7} + (I+L^{24})(I+R^7)v_{n-8}
\end{split}
\end{equation}

where $L$ is a left shift matrix and $R$ a right shift matrix and the exponent
the number of bits being shifted.

\subsubsection{\label{seckiss} KISS}

The KISS (\emph{keep it simple stupid}) has been proposed by Marsaglia as the
combination of three RNGs: multiply-with-carry, xorshift and
congruential~\cite{PARK88}. The first two RNGs have already been discussed in
previous sections. The XorShift generator is however different from the one
that has already been presented. Indeed, the original XorShift generator
proposed by Marsaglia will be used~\cite{MARS03}. The proposed generator is a
type I XorShift generator~\cite{PANN09}. These generators have $r=1$, meaning
that the state is represented by only  one $w=32$ bits variable, and $A_1$ is
the product of three xorshift matrices or, in other words, $s=3$. The following
$A_1$ matrix will be used, as proposed by Marsaglia:
\begin{equation}
A_1 = (I+L^{13})(I+R^{17})(I+L^5)
\end{equation}
The above XorShift generator was shown by Panneton~\cite{PANN09} to fail
randomness tests of the TestU01 suite by L'Écuyer~\emph{et al.}~\cite{LECU07}.
However, as will be explained later, combining several random number generators
can have a positive effect on the randomness as well as the period of the
resulting generator.

The new RNG introduced in this section as part of the combined RNG, the linear
congruential generator (LCG)~\cite{PARK88}, is similar to the MWC generator.
The recurrence has the following form:
\begin{equation}
X_{n+1} = (aX_n+c) mod (m)
\end{equation}

where $a$ is a multiplier, $c$ the increment (as opposed to the carry in the
MWC) and $m$ the modulus. Again, the choice of $a$ is not left to chance. In
order to obtain the full period $p = m -1$, the modulo $a$ must be prime and
$a$ must be a primitive element modulo $m$~\cite{LECU99}. However, since the
objective of this work is the implementation of efficient RNGs, we will forgo
the full period LCG and opt for an efficient modulo of $2^{32}$, which is
essentially free through integer overflow. Marsaglia proposes the values
$a=69069$, $c=1234567$ and $m=2^{32}$~\cite{MARS03}.

The final random number value is obtained by xoring the MWC and LCG
intermediate values to which the XorShift value is added.

\subsubsection{Parallelizing RNGs}
If RNGs are not suitably parallelized, running them on a parallel architecture could result in every
substream outputing the same exact sequence of pseudo-random numbers, thus rendering useless
the parallel architecture or its resulting correlated streams. For example, if only different seeds are sent
to every node, the seed sent to node \emph{j} could be the second random number generated by node
\emph{i}, making substreams \emph{i} and \emph{j} identical but lagged by one.

Two main categories of parallelization can be used for generating
non-correlated streams on a parallel architecture:  those that split a given
generator into multiple substreams and those that generate multiple independent
streams~\cite{BADA06}. From the first category, we again find two
subcategories: sequence splitting and leapfrogging.

In sequence splitting, a processor or node \emph{j} will generate the random
numbers $[R_j, R_{j+1}..R_{j+B}]$ where $B$ is the amount of random numbers
generated by each node and $R_i$ is a random number. It can easily be seen that
this can cause problems if it is not known beforehand how many numbers a given
node will need to generate and if this amount happens to be larger than B. This
technique can be employed with RNGs capable of computing the \emph{i}'th random
number generated without computing all the intermediate numbers. From this
property, every node \emph{j} can be seeded with the last value that would have
been generated by node \emph{j-1}. The generator itself does not have to be
modified.

In leapfrogging, node \emph{j} generates the random numbers $[{R_j,R_{j+N},R_{j+2N} ,...}]$ where $N$ is the number of processors executing the RNG. It is clear that
in this case the sequences will not overlap. This technique also requires a way to
compute the random number \emph{i} without having to compute all the intermediate values. In this case,
the RNG has to be modified in order to skip N numbers every time it has to generate one number.

Finally, the second category of RNG parallelization, called parameterizing, achieves uncorrelated
sequences by effectively using a different RNG for all nodes. These different RNGs can be of the same
type, \emph{v.g.} a MWC generator, but with different generator parameters, \emph{e.g.} with different
multipliers \emph{m}.

\subsection{Graphics cards for scientific computing}
The recent interest for graphics card computing within the scientific community
stems from the fact that GPUs (Graphics Processing Units) offer a low cost
alternative to traditional CPU clusters for parallel computations. Graphics
hardware being inherently parallel, it is useful for a large array of
scientific computations such as dense linear algebra~\cite{GARL08}, molecular
dynamics~\cite{ANDE08}, N-Body simulations~\cite{NYLA07}, and so forth.

This work relies on the NVIDIA CUDA~\cite{NVIDIA07} platform which allows the
programming of graphics cards using a syntax that is heavily based on the C
language. The CUDA platform is available for all graphics card of the G80
family and beyond. The Unified Shader Model was introduced in this family,
which effectively transformed the graphics card into a massively parallel SIMD
(single instruction multiple data) or, as NVIDIA puts it, SIMT (single
instruction multiple threads) machine where several hundred of identical
computing elements, Scalar Processors (SPs) in NVIDIA terminology, are present.

The MultiProcessor (MP), a group of 8 SPs, is equipped with 16KB of shared
memory, a fast cache that is available to all SPs of a MP. All the SPs also
have access to the large global memory pool (256~MB and up). The global memory
however presents bandwidth and latencies that are vastly inferior to shared
memory. Current GPUs are equipped with 1 to 30 (or 60 in a dual-GPU graphics
card) MPs for a total of up to 240 (or 480) SPs. Two Special Function Units
(SFUs) are also present, capable of executing transcendental functions or a
floating point multiply operation.

The global CUDA paradigm consists in launching thousands of threads, grouped in \emph{thread
blocks}. These blocks will be assigned to a given MP. A given MP can be assigned to multiple thread
blocks that could be scheduled to launch concurrently (in warp locksteps) if the resources of the MP
allow for more than one block to be launched, \emph{v.g.} if there are enough shared memory and
registers to accommodate both blocks.

The smallest computational group is called the \emph{warp} and in the GT200 architecture is composed of
32 threads. An SM will select a warp that is ready to execute and every thread of the warp will execute
the same instruction.

Integer multiplications are considered slow on NVIDIA graphics card. Indeed,
the integer multiplication of two 32 bits operands has a throughput of 2
operations per clock cycle whereas the throughput of the single precision
floating point multiplication has a throughput of 8 operations per clock cycle.
This could have an important impact on the performance of the MWC RNG. Other
integer operations useful for the scope of this paper, such as bitwise XORs,
shifts and additions have a throughput of 8 instructions per clock cycle. The
GPU also provides a lower precision 24 bit integer multiplication with a 32
bits results and an 8 instructions per clock cycle throughput. However, NVIDIA
announced that in future hardware versions, the lower precision multiplication
will become slower than the full 32 bit multiplication.

\section{Methods}

\subsection{\label{secgpuimp} GPU implementations}

This section will cover the implementation details for the different RNGs that
have been used. It is however necessary to explain why the Mersenne Twister
(MT)~\cite{MATS98} has not been implemented, which is the RNG of choice when an
extremely long period is required. The MT RNG requires a fairly large state
vector (624 integers, 2496 bytes) in order to generate random numbers. This
state has to be stored somewhere and access to fast enough memory is not
sufficient. Consequently, the state vector would need to be stored in global
memory on the GPU, therefore severely hindering performances. Indeed, since the
thread blocks will usually contain more than 32 threads and that each thread
needs its own state vector, there is not enough registers or shared memory
space to store the state vectors of a thread block. It has therefore been
decided to use RNGs with relatively small state vectors at the cost of a
reduced period. The NVIDIA CUDA SDK already has an implementation of the
MT19937 random number generator which will be included in the comparison to our
results. NVIDIA has used global memory to store the large state vector.

\subsubsection{multiply-with-carry}

In the current implementation, two MWCs are combined together as has been
originally proposed by Marsaglia~\cite{MARS97}. There is evidence~\cite{LECU88}
that combining two or more RNGs not only increases the period but also the
randomness of the sequence. These two MWCs therefore need independent
multipliers. Since the 32-bit arithmetic and logical units (ALU) of the GPU are
used, the multiplier $a$ and carry $c$  values have to fit inside 16 bits and
the base $b$ is set to $2^{16}$. There is a limited number of choices (392 to
be exact) for the multiplier $a$ that satisfy the conditions exposed
in~\ref{ssmwc}. This is seemingly not enough for the thousands of independent
RNGs being launched, in accordance with the CUDA paradigm. However, it is the
combined results of two MWCs to generate one sequence of random number per
thread that is of interest. It is therefore the number of distinct combination
of two multipliers that needs to be sufficient, and that number (k-combination
of 2 in 392) is plentiful.

The MWC generator is therefore parallelized through parametrization where
independent streams are generated by each RNG (or thread). A different set of
seeds (two) is also given to every thread. Each thread generates a number of
random numbers (>1000) before exiting.

The execution flow for the CUDA kernel starts by having each thread fetch its corresponding seeds
and multipliers values from global memory. These memory transactions are completely coalesced. A
loop having as many iterations as the numbers of random numbers to be generated by every thread
then begins. The core of the loop generates two intermediate numbers, one for every intermediate
MWC. These two 16 bits intermediate values are concatenated to form the final 32 bits integer random number. This
number is written to global memory.

The total number of useful operations for the core of the loop to generate one
random number is 10 (3 $shifts$, 2 $ands$, 2 $mults$ and 3 $adds$). This is
confirmed by inspection of the generated assembly code, with the addition of
address calculations for the storage of the random number and loop overhead.

Fourteen registers are used by this implementation. Using the NVIDIA occupancy
calculator, the optimal block sizes of 128, 256 or 512 are found, which give an occupancy of 1.0 on the
NVIDIA GTX280 card.

\subsubsection{XorShift}

The XorShift generator used in this work is the one proposed by
Panneton~\cite{PANN09} and presented in Section~\ref{secxorshift}.

Using values of $r=8$ and $w=32$, a state vector of eight 32 bit values has to
be stored for every generator or, in other words, every thread. Ideally, this
state vector is to be stored in registers. However, as of CUDA 2.3, the
registers are not indexable, \emph{v.g.} the $[ ]$ operator cannot be used,
which is required by the algorithm~\cite{PANN09} to address the state vector.
Shared memory, which is addressable, has therefore been used, as a first
approach to store the state vector. Each thread then needs $r*w=256$ bits of
shared memory.

\begin{figure}[htp]
\begin{center}
\includegraphics[width=\linewidth]{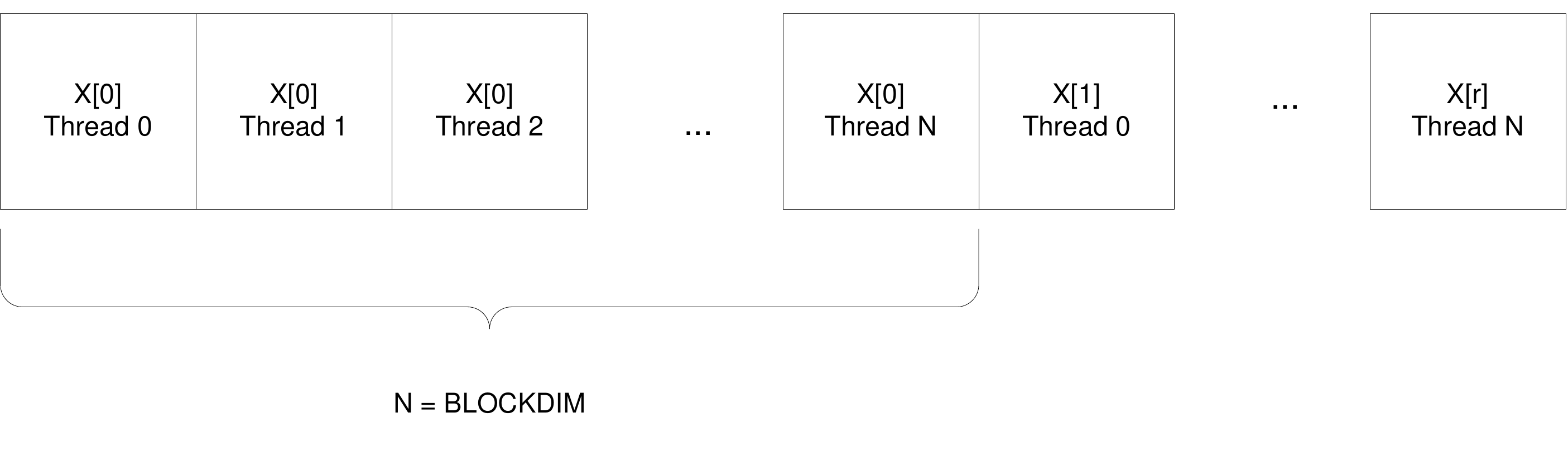}
\end{center}
\caption{\label{figshared}Shared memory arrangement for the XorShift RNG.}
\end{figure}

To fully exploit the bandwidth of the on-chip shared memory, certain guidelines
have to be followed~\cite{NVIDIA07}. Shared memory is divided into banks that
can be accessed simultaneously. If $N$ requests fall into $N$ different memory
banks, these requests can then be serviced simultaneously. In the GT200
architecture used for this work, shared memory is divided into 16 banks (equal
to the number of threads in a half-warp) and successive 32 bits words are in
successive banks. It is therefore important to allocate the vector in shared
memory in a way that will not cause bank conflicts, where two requests would be
sent to the same memory bank. The scheme (explain the scheme shortly) presented
in Figure~\ref{figshared}, which assigns successive state vector element for a
given thread with a $blockDim$ stride (where $blockDim$ is the size of the CUDA
thread block), has been used resulting in a bank conflict free implementation.

The rest of the XorShift algorithm maps natively to the programming model in CUDA as only $shift$ and
$XOR$ operations are needed. The execution flow for the CUDA kernel starts with each thread fetching
its correspondent 8 seeds from global memory and storing them to the shared memory. Again, these
memory transactions are completely coalesced. A loop having as many iterations as the number of
random numbers to be generated by every thread then begins. The core of the loop operates on two
intermediate values $t$ and $y$. These $t$ and $y$ values are modified through the recurrence
presented in Section~\ref{secxorshift}. The resulting $y$ value is written back to the state vector in
shared memory as well as to the results array in global memory.

The number of useful operations to generate one random number is 18 (11 $XORs$
and 7 $shifts$). The C implementation proposed by Panneton~\cite{PANN09}
outputs one random number every time it is called. It therefore needs to
implement an implicit circular buffer over the $r$ elements of the state vector
and increment the current vector element pointer, $k$, every time the routine
is called. Consequently, the address and index calculations must be determined
at runtime for every generated random number since the indices depend on $k$.
Logical $and$ operations also have to be used to implement a circular buffer
over the $r$ elements. This adds many operations that are not directly useful
to the generation of random numbers. It results that 52 assembly operations are
required to generate one random number.

We have no doubt that Panneton~\cite{PANN09} proposed the implementation as an ``inline'' random
generator where one random number is requested for every call. It has
been implemented as is as an experiment. However, since the goal of this work is to generate a large
quantity of random numbers, we can implement the circular buffer explicitly with $r$ consecutive
statements in a manner similar to loop unrolling.  With this technique, $r$ random numbers are
generated for every execution of the main loop. The
total number of assembly instructions is brought down to 23 per random number.

At this point, there is no need for an indexable memory space since the
circular buffer has been made explicit. It is therefore possible to replace the
shared memory array of $r$ elements with $r$ registers. This decision should be
made by taking into consideration which resource (registers or shared memory)
will be most likely limited when the RNG is sampled ``on-the-fly'' by a
complete simulation engine also requiring resources. By using registers instead
of shared memory, the total number of assembly instructions per random number
is found to be an optimal 18.


The number of registers used by this implementation is 14 for the explicit loop and 15 with
the unrolled implementation. Using the NVIDIA occupancy
calculator, the optimal block sizes of 128, 256 or 512 are found, which give an occupancy of 1.0 on the
NVIDIA GTX280 card.

\subsubsection{KISS}

The implementation of the KISS generator relies on the previously discussed implementation of
the MWC generator, together with a simplified XorShift generator and the congruential generator.

The simplified XorShift generator has $r=1$, $w=32$ and only one unsigned integer is therefore
needed to store the state vector. The use of shared memory is consequently avoided. The newly
introduced LCG RNG is implemented with a single line of code that can be easily derived from the
recurrence presented in Section~\ref{seckiss}.

The same parallelization schemes have been used for the two previously discussed RNGs: different
multipliers as well as different seeds for the MWC generator and different seeds to the XorShift
generator. The LCG is parametrized using the multiple seeds scheme as well.

The state vector for the combined generator is therefore composed of four 32 bits words (2 for  the MWC,
1 for the simplified XorShift and 1 for the LCG). In addition to these seeds, the two multipliers of the
MWC also have to be fetched from global memory. The rest of the implementation is similar to what has
previously been mentioned in the two previous sections.

The number of useful operations for the main loop of combined generator is
found to be 18. In terms of useful operations it is therefore equivalent to the
XorShift RNG. However, the need for shared memory has been completely avoided
which translates in a reduction of the number of overhead instructions
associated with the memory address calculation when compared to the implicit
implementation of the XorShift generator. This is apparent in the assembly code
file where fewer lines of code are generated for the KISS RNG compared to the
implicit XorShift RNG. The newly introduced LCG RNG uses a 32 bit integer
multiplication which has a throughput four times lower than the other integer
operations used.

\subsection{Performance evaluation}
For all implementations, approximately 80 millions random numbers are generated by each call to the
random number generator. The random numbers are written to global memory. They could therefore
be used by subsequent CUDA kernels or copied back to the computer's main memory for use by the
CPU. The execution times also include the writing of the state vector to global memory for subsequent
use. The timing results do not include the copying
 of the results back to the system's main memory. The rate at
which the random numbers are generated, in samples per second, will be reported.

Furthermore, the rate at which the random numbers are generated when there is
no need to write them back to global memory has also been studied. This result
is of interest for applications that would inline the random number generator
instead of generating a pool of random numbers for later use. To achieve this
result, intermediate results are not written to global memory until the end of
the main loop, at which point the state of the RNG is written to memory to
ensure that the whole loop is not simply suppressed by optimization during
compilation.

The achieved memory bandwidth ($BW$) and usable operations per seconds  will
also be presented with regards to the theoretical limit of the hardware used.
The memory bandwidth is measured as:
\begin{equation}
BW = \frac{n_t*N_t*4}{t*10^{9}} GBps
\end{equation}
where $n_t$ is the number of random numbers generated per thread, $N_t$ the
number of threads invoked, $t$ the execution time and the $4$ factor is used to
account for the size $w=32 bits=4 bytes$ of the generated number. It will only
be of interest when the results are written back to global memory. The number
of usable operations per second ($U_{ops}$) result is measured as:
\begin{equation}
U_{ops} = \frac{n_t*N_t*n_{ops}}{t*10^{9}} GOps
\end{equation}
where $n_{ops}$ is the number of operations required to generated a random number as presented in
Section~\ref{secgpuimp}. The simpler FLOPs (floating point operation per second) measure cannot be
used here since no floating point operations are taking place except when the generated number is
scaled back to [0,1[ range.

Two types of random numbers generators have also been implemented, resulting in
two types of random numbers: unsigned integers in the [0,$2^{32}$-1] range and
single-precision floating-point numbers in the [0,1[ range. The latter requires
two extra steps to convert the integer number to a floating point
representation and to scale it to the [0,1[ range through a single-precision
floating-point multiplication.

The main loop, which generates a series of random numbers within a kernel, has been unrolled using
the supplied $pragma$ to the $nvcc$ compiler in order to reduce the loop control logic overhead. The
compiler has chosen not to unroll the main loop of the XorShift generator with the explicit circular buffer most likely due to register pressure.

Finally, the randomness of the implemented RNGs has been assessed with the use
of the DIEHARD~\cite{MARS95} and TestU01~\cite{LECU07} applications. TestU01
supplies three levels of randomness testing: \emph{SmallCrush, Crush} and
\emph{BigCrush}. The \emph{Crush} battery of tests requires approximately
$2^{35}$ random numbers to complete. The DIEHARD battery of test is less
stringent, as is apparent by the relative time it takes to complete when
compared to TestU01: around 10 seconds compared to one hour. The quality of the
three random number generators has been studied, first by the use of the
DIEHARD battery of test, then by the \emph{Crush} battery of tests from
TestU01.

The results have been obtained with an NVIDIA GTX280 graphics card with 1GB of global memory
and on a Xeon Q6600 processor. The CUDA $kernels$ have been compiled with
the $nvcc$ compiler included in the version 2.3 of the toolkit and the CPU programs with the -O2
options activated using the compiler included with Microsoft Visual Studio 2005.

\section{Results}

In the following tables, the results are presented for the unsigned integer random generator.
The first column indicates if the results are written back to memory. The last column of the tables,
$uniform$ $rate$, presents the rate for a uniformly distributed random number generator.

Tables~\ref{tab:resmwc},~\ref{tab:resxorshift} and~\ref{tab:reskiss} present the results for the GPU
implementation of the MWC, XorShift and KISS RNGs, respectively.

\begin{table}[htp]
\begin{center}
\begin{tabular}{lccccc}
\hline
   & $t$  & $BW$   & $U_{ops}$ & $rate$ & $uniform$ $rate$\\
   & $(ms)$  & $GBps$   & $ GOps $ & $GSamples/s$ & $GSamples/s$ \\
\hline  \hline
W/ writeback & 2.3 & 67.2& 168.1 & 18.0 & 17.9 \\
W/o writeback & 1.9 & - &  300.6 &37.5& 23.6\\
\hline
\end{tabular}
\caption{\label{tab:resmwc} Results for the MWC generator.}
\end{center}
\end{table}

\begin{table}[htp]
\begin{center}
\begin{tabular}{llccccc}
\hline
&   & $t$  & $BW$   & $U_{ops}$ & $rate$ & uniform rate\\
  & & $(ms)$  & $GBps$   & $ GOps$ & $GSamples/s$ & $GSamples/s$ \\
\hline  \hline
Implicit &W/ writeback & 9.1&	16.6	&74.9	&4.4	&4.3 \\
& W/o writeback & 8.5	&-	&81.1&	4.8&	4.4\\
\hline
Explicit &W/ writeback & 5.1& 30.2& 145.7& 8.1 & 7.6\\
&W/o writeback & 3.0& - &248.2 &13.8 &11.5\\
\hline
Registers &W/ writeback &3.3 & 46.4& 224.1& 12.5&11.5\\
&W/o writeback & 2.8& - & 264.5& 14.5&12.9\\
\hline\hline
\end{tabular}
\caption{\label{tab:resxorshift} Results for the XorShift generator.}
\end{center}
\end{table}

\begin{table}[htp]
\begin{center}
\begin{tabular}{lccccc}
\hline
   & $t$  & $BW$   & $U_{ops}$ & $rate$ & uniform rate\\
   & $(ms)$  & $GBps$   & $ GOps $ & $GSamples/s$ & $GSamples/s$ \\
\hline  \hline
W/ writeback & 4.4	&34.2&	153.9&	9.2	&8.5 \\
W/o writeback & 2.6&	-&	304.8	&19.1&	10.3\\
\hline
\end{tabular}
\caption{\label{tab:reskiss} Results for the KISS generator.}
\end{center}
\end{table}

\subsection{Randomness}

The randomness of our implementation of these RNGs has been studied with DIEHARD and the \emph{Crush} battery of tests
from TestU01. All random generators have passed all tests from the DIEHARD battery of tests. The
KISS generator has passed all tests from the \emph{Crush} battery of tests while both the XorShift and
MWC have failed exactly 1 test, out of the 96 applied, from the \emph{Crush} battery of tests.

\subsection{Comparison with other studies}

Table~\ref{tab:rescomp} presents a comparison between our GPU implementations,
the equivalent single threaded CPU implementation as well as a collection of
random number generators found in the literature. The comparisons to other GPU
work has to take into consideration the fact that different GPUs have been used
to generate the results. Thomas \emph{ et al.}~\cite{THOM09} have used an
NVIDIA GTX280, Langdon~\cite{LANG09} a NVIDIA 8800GTX and Janowczyk \emph{et.
al}~\cite{JANO08} a GeForce 8800GTS. It should also be noted that the XorShift
implementation presented by Thomas \emph{ et al.} is not equivalent to ours: it
uses $r=5$ and the number of $xorshift$ operations, $s$, is unknown. Their
implementation thus uses less memory for the state vector than ours as well as
possibly requiring less instructions to generate one random number.


A study of the efficiency of the implementations, in terms of MSamples/Joule, has been performed as
well as a cost-benefit study in terms of MSamples/s/\$. In order to acquire the power consumption of
every devices in our comparison, the TDP (thermal design power) value has been used. An online
retailer's prices have been used with the exception of the GeForce 8800GTX and 8800GTS which are
no longer available for sale.


\begin{table}[htp]
\begin{center}
\begin{tabular}{l||ccc}
\hline
   & Efficiency & Cost-Benefit & Throughput\\
   & $10^6$ Samples/Joule & $10^6$ Samples/s/\$ & $10^9$ Samples/s\\
\hline  \hline
MWC   &   76.3 & 78.3  & 17.9 \\
XorShift &    52.9 & 54.3   &12.5 \\
KISS    &   36.0&  36.7  & 8.5 \\
$MWC_{CPU}$                     &    2.8 &  1.4    & 0.3 \\
$XorShift_{CPU}$                &    1.0 & 0.5    & 0.1 \\
$KISS_{CPU}$                    &    1.9 &  0.91    & 0.2 \\
\hline
$MT19937_{FPGA}$\footnote[1]{Actual FPGA used not mentioned in
the text}~\cite{VINA07} &    - & -
& 119.6 \\
\hline
$SFMT_{CPU}$\footnote[2]{SSE Optimized} ~\cite{THOM09} &    32.7 &  13.6     & 4.3 \\
$XorShift_{GPU}$~\cite{THOM09} &  71.5 &  73.4    & 16.9 \\
$Lut-Opt_{FPGA}$~\cite{THOM09} &  8636.7 &  108.2     &259.1\\
\hline
$Park-Miller_{GPU}$~\cite{LANG09} &  5.4 &  -       &$\approx 1$\\
\hline
$Lagged Fibonacci_{GPU}$~\cite{JANO08} & 10.7  & - & 1.3\\
\hline
$Mersenne Twister_{GPU}$~\cite{PODL07} & 12.7& 13.0 & 3.0\\
\end{tabular}
\caption{\label{tab:rescomp} Comparative results.}
\end{center}
\end{table}



\section{Discussion}

Tables~\ref{tab:resmwc},~\ref{tab:resxorshift} and~\ref{tab:reskiss} provide an
overview of the different RNGs implemented in this work. The GTX280 has a
theoretical FLOPs value of 933. The figure assumes that each SP is capable of
executing 3 FLOPs per cycle (one fused multiply-add on the SP and one multiply
on the SFU). Very special circumstances have to be met to sustain the
theoretical FLOPs figure. In the present case, no fused multiply-add will take
place and no operation will be sent to the SFU since only integer math is used.
The maximal achievable operations per second figure in our case is therefore
one third of the maximum theoretical FLOPs figure. To this inherent limitation,
memory calculation arithmetic as well as loop overhead have to be acknowledged
when looking at the achieved operations per second figure. The theoretical
memory bandwidth of the NVIDIA GTX280 is 141.7 GBps. Our own benchmark has
revealed that the achievable bandwidth is closer to 117GBps.

The $U_{ops}$ value for the two RNGs not requiring shared memory, the MWC and the KISS
generators, show good agreement with the above discussion. They respectively achieve 96.7\% and
97.8\% of the maximum theoretical operations per second.

In the XorShift RNG with the implicit circular buffer, the overhead of
 index and address calculations become apparent. Indeed, 6 instructions are required to fetch one
piece of data from shared memory. The overhead would be the same if registers
were indexable so it is not a question of shared memory versus indexable
registers. In the end, more than half of the operations are required by the
index calculation and implementation of the circular buffer, and thus not
directly used for random number generation, which explains the low $U_{ops}$
value of 26\%. With the explicit version of the same RNG yielding the same
results, the observed $U_{ops}$ value raises to 79\% of the maximum theoretical
operations per second. When avoiding shared memory completely and using only
registers to store the state vector, this figure raises to 85\%.

It can also be seen that the cost of converting the integer value to the [0,1[ range is almost completely
hidden by the bandwidth limitation as there is almost no difference between the integer and uniform
samples/s rate when the results are written back to memory. However,  when the RNGs are
used ``on-the-fly'', requiring no writes to global memory, the conversion and subsequent normalization
are responsible for a performance drop of almost 50\% in the worst case.

The achieved memory bandwidth does not go over 50\% of the observed achievable bandwidth. Our explanation is that this behavior is related to the fact that only memory writes take place during the main loop of the $kernels$. The same results are obtained when modifying the benchmark used to find the achievable bandwidth so that only memory writes occur.


\subsection{Comparison with other studies}

It is clear that GPUs are an attractive platform for the generation of random numbers, as this study and other studies have shown.

The proposed MWC generator is the fastest GPU-based RNG found in the literature. Its
implementation falls natively within the GPUs strength, using only the fastest operations available on
the GPU will requiring a small vector state. The proposed XorShift generated is slower than the
XorShift generator proposed by Thomas \emph{et al.} but, as it was said, we are unsure of the actual number of operations required by their implementation to generate one random number, which directly impacts the $rate$ value. We have followed Panneton's recommendation of using more than three XorShifts
operations on the state vector to achieve acceptable randomness and we are again unsure if their implementation respects this.

The FPGA platform offers a seemingly unparalleled solution to random number
generation. One aspect to keep in mind is that the FPGA implementations
presented here most likely fully use the configurable hardware present on the
FPGA to generate the reported number of samples per second. In a real world
situation, the generated random numbers have to serve a purpose. For example,
they could be used within a Monte Carlo simulation. The FPGA RNG
implementations would therefore have to allocate a significant  portion of the
configurable hardware to the simulation logic, reducing their random number
production rate accordingly. The generated numbers could also be transferred
through one of the FPGA's extension card to some other hardware platform
responsible for the simulation, consequently being limited by the connection
between the two devices and once again sacrificing the effective production
rate. The higher cost of FPGAs also explains a much smaller gap in the
cost-benefit column of Table~\ref{tab:rescomp} between FPGAs and GPUs.

\section{Conclusion}
This paper has presented the GPU implementation and parallelization of three
random number generators: the multiply-with-carry, the XorShift and the KISS.
For each generator, the exploitation of the GPU's resources have been
evaluated. The results indicate that the GPU is an attractive platform for
generating random numbers. It has also been shown that the intensive use of
shared memory as an indexable storage space comes with a high address
calculation cost and a significant loss of useful operations per second
performance.

\section{Acknowledgements}
This work has been supported by the Natural Sciences and Engineering Research
Council of Canada CRSNG/NSERC.

\section*{References}
\bibliographystyle{myunsrt}
\bibliography{biblio}

\end{document}